\documentclass[10pt,letterpaper,twocolumn]{article} %
\usepackage{ol}
\usepackage{hyperref}
\usepackage{amsmath}

\begin{document}
\twocolumn[

\title{Scanning Near-Field Optical Coherent Spectroscopy of\\ Single Molecules at 1.4 Kelvin}

\author{Ilja Gerhardt, Gert Wrigge, Mario Agio, Pavel Bushev$^*$, Gert Zumofen, and Vahid Sandoghdar}
\affiliation{Laboratory of
Physical Chemistry, ETH Zurich, CH-8093 Zurich, Switzerland}

\begin{abstract}We present scanning near-field extinction
spectra of single molecules embedded in a solid matrix. By varying
the molecule-tip separation, we modify the line shape of the
spectra, demonstrating the coherent nature of the interaction
between the incident laser light and the excited state of the
molecule. We compare the measured data with the outcome of numerical
calculations and find a very good agreement.
\end{abstract}

\ocis{270.1670, 180.5810, 290.2200, 300.6250}

]

\noindent Scanning Near-field Optical Microscopy (SNOM) was invented
about twenty years ago and has been used in a wide range of
applications where a subwavelength spatial resolution is
advantageous in optical studies\cite{paesler_book}. One of the
highlights of SNOM was to deliver the first images of single
fluorescent molecules at room temperature in 1993\cite{betzig07},
initiating the active and fruitful field of single molecule
microscopy. Since then various groups have combined SNOM with
spectroscopy of single
emitters\cite{hess01,moerner03,matsuda01,guest01} by detecting the
inelastic fluorescence that is red-shifted with respect to the
excitation light. Developments in both
far-field\cite{plakhotnik12,lindfors01,hoegele01} and
near-field\cite{mikhailovsky01} spectroscopy of single nano-objects
have shown, however, that it is possible to study them via the
interference between the elastically scattered light and a reference
beam. Very recently, we used this approach to demonstrate the first
near-field extinction measurement on single molecules and reported a
dip in transmission as large as $6\%$ without using any noise
suppression method\cite{gerhardt03}. In this Letter, we present the
dependence of the single molecule extinction spectra on the
molecule-tip separation and compare our findings with the outcome of
numerical simulations.

\begin{figure}[htb]
\centerline{\includegraphics[width=7.5cm]{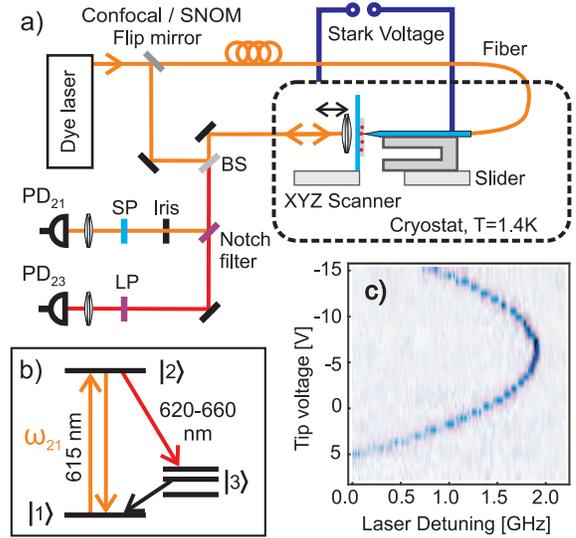}}
\caption{\label{setup}a) Schematics of the experimental setup. BS:
beam splitter, SP (LP): short (long) pass filter. b) The
level-scheme of a dye molecule. c) The Stark-shift recorded as a
function of the voltage applied to the tip.}
\end{figure}
\normalsize

The experimental setup is schematically depicted in Fig.~\ref{setup}
and consists of a combined scanning confocal and near-field optical
microscope that operates at T=1.4~K. Details of this setup, the tip,
the sample and the theoretical concepts of our work have been
described in Ref.\onlinecite{gerhardt03}. In a typical experiment,
we first detected single molecules via near-field fluorescence
excitation spectroscopy\cite{moerner03}, i.e. by recording the
Stokes shifted emission on transition $2\rightarrow 3$ in
Fig.~\ref{setup}b). We monitored the oscillation of a quartz tuning
fork and used the shear-force interaction to control the tip-sample
distance\cite{karrai01}. In this manner, we could position the tip
at a given axial distance from the sample (thickness $\sim 50-100~nm
$) to within 10~nm. Upon lateral scanning of the tip, we found that
despite electrical contacting of the tip to the ground, a
substantial position dependent Stark shift was consistently present
in all experiments. In order to compensate this effect, we first
laterally centered the tip on the molecule by maximizing the
fluorescence signal on detector $PD_{23}$. We then varied the
voltage on the SNOM tip to the value corresponding to the apex of
the voltage-shift parabola, as shown in Fig.~\ref{setup}c. This
procedure allowed a near to full cancellation of the Stark shift,
indicating that its origin is probably charge build-up on the sample
or the oxidized surface of the tip.

\begin{figure}[htb]
\centerline{\includegraphics[width=8.4cm]{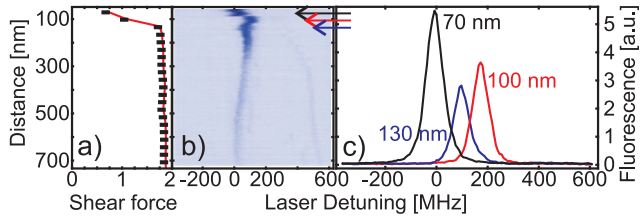}}
 \caption{\label{broadening}Shear-force amplitude (a) and simultaneously recorded fluorescence excitation spectra (b) as a function of tip-sample separation. c) Three exemplary spectra from the indicated tip-sample distances.}
\end{figure}

Even in the absence of an external electric potential, the presence
of the tip could affect the linewidth and position of the molecular
resonance~\cite{bian01} similar to the case of a molecule in front
of a mirror. Our three-dimensional finite-difference time-domain
(FDTD) calculations show that for a dipolar resonance with linewidth
$\gamma_0$, we should expect a broadening in the range of $\gamma_0$
to $3\gamma_0$ (depending on the molecular dipole orientation) for a
tip-molecule separation of 30~nm. In our experiment, it turned out
that we could not probe these effects because as shown by an example
in Figs.~\ref{broadening}b and c, we often encountered spectral
instabilities for tip-sample separations under about 100~nm. The
correspondence between the data in Fig.~\ref{broadening}b and the
simultaneously recorded shear force signal displayed in
Fig.~\ref{broadening}a indicates that this effect is due to a
mechanical perturbation of the \emph{p}-terphenyl matrix and that
moreover, our shear-force signal is assisted by mechanical
contact~\cite{gregor01}. Although this phenomenon merits further
investigations, in this work we have chosen to avoid it by operating
the tip at distances larger than about 60~nm from the sample where
the tip influence is negligible.

\small
\normalsize

\begin{figure}[htb]
\centerline{\includegraphics[width=7.5cm]{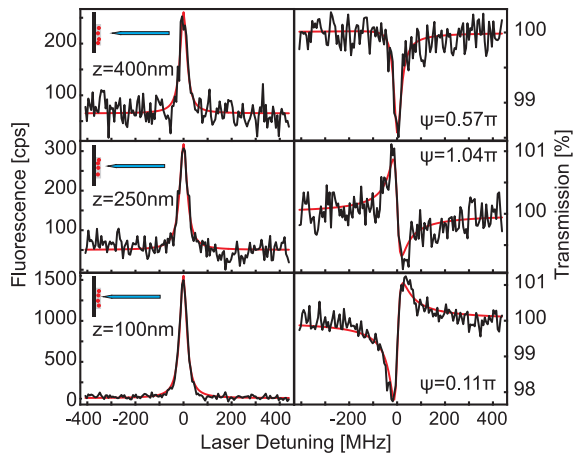}}
 \caption{\label{exdip}Simultaneously recorded fluorescence signal ($PD_{23}$) and
extinction measurement ($PD_{21}$) for three different heights from
the tip to the sample. Solid curves show fits as discussed in Ref.
\onlinecite{gerhardt03}}
\end{figure}

The signals on the two detectors $PD_{23}$ and $PD_{21}$ have been
described in Ref.\onlinecite{gerhardt03} and its Supplementary
Material. In short, while $PD_{23}$ records the conventional Stokes
shifted fluorescence $I_{23}$, the signal on $PD_{21}$ is the result
of the interference between the transmitted laser light through the
tip and the light that is coherently scattered by the molecule.
Thus, a change in the position of the tip with respect to the
molecule can lead to a change in the relative accumulated phase and
a change in the resonance shape. Figure~\ref{exdip} shows examples
of such spectra at three different tip-molecule positions.

\small
\normalsize

The coherent spectra recorded on $PD_{21}$ can be fitted by
$I_{bg}-I_{bg}V\frac{\left( \Delta \cos \psi + \frac{\mathit{\gamma}
}{2}\sin \psi \right)}{ \Delta ^{2}+\gamma ^{2}/4}$ if the intensity
of the molecular emission is much smaller than the strength of its
interference with the laser field, as is the case in our
experiment\cite{gerhardt03,plakhotnik12}. Here $\Delta$ denotes the
detuning between $\omega_{21}$ and the excitation laser frequency,
$\gamma$ is the transition linewidth, $\psi$ stands for the relative
phase between the excitation and scattered lights, and $V$ is a
parameter that we call visibility. We remark that in these
experiments an iris of diameter 1~mm was used in the path of
$PD_{21}$ to select the axial part of the beam.

\begin{figure}[hbt]
\centerline{\includegraphics[width=7.5cm]{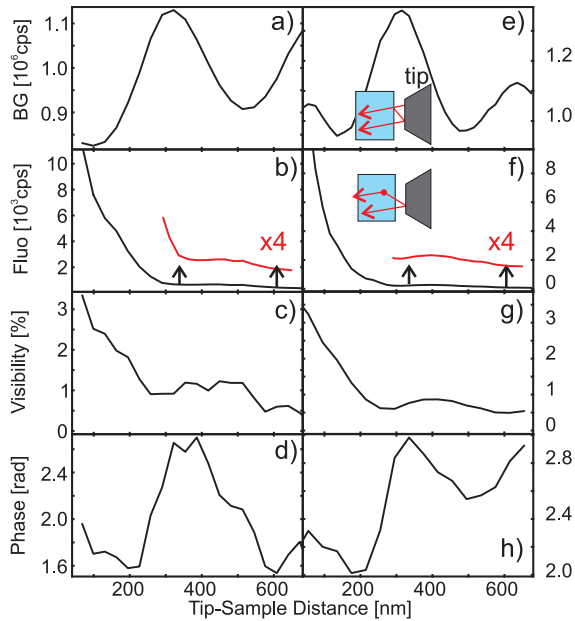}}
 \caption{\label{fdtd}Experimental (a-d) and FDTD (e-h) data for the the nonresonant transmission (BG),
the fluorescence signal $I_{23}$ (Fluo), the visibility $V$ and
phase $\psi$ of the extinction signal on $PD_{21}$ as a function of
the separation between the sample surface and a tip of 200 nm
aperture.}
\end{figure}

Figures~\ref{fdtd}a-d summarize the analysis of a series of spectra
from different axial tip-molecule separations for a different tip
and molecule than in Fig.~\ref{exdip}. In order to compare our
experimental data with theoretical expectations, we have performed
three-dimensional FDTD calculations. The 50~nm-thick matrix was
modelled according to the refractive indices of \emph{p}-terphenyl
along its crystal axes, and the molecule was taken to be a
dispersive Lorentz material along its dipole moment. The tip was
taken to be 700~nm long and buried 50~nm into convolutional
perfectly-matched-layer absorbing boundary conditions to avoid
finite-size effects\cite{roden}. The total electric field of the tip
and the molecule was recorded on a reference sphere of radius
$1.2~\mu m$ centered at the molecule position. We verified that this
field is transverse to within $99\%$. The frequency dependent
signals obtained were fitted in the same fashion as the experimental
data.

Figure~\ref{fdtd}a plots the excitation intensity $I_{bg}$
determined from the off-resonant tail of each frequency scan
recorded on $PD_{21}$ while Fig.~\ref{fdtd}e shows the corresponding
FDTD results. As sketched by the inset in Fig.~\ref{fdtd}e, the
oscillatory behavior of $I_{bg}$ is caused by the interference
between the part of the laser light that exits the tip and directly
propagates to the detector and a second part that is first reflected
from the sample and then from the tip (i.e. twice $\pi$ reflection
phase shifts) before traversing the detection path. This oscillatory
behavior has been also reported previously in conventional SNOM
experiments\cite{hecht15}. The FDTD calculations also clearly
reproduce this effect.

Figure~\ref{fdtd}b displays $I_{23}$ extracted from Lorentzian fits
to the fluorescence excitation spectra (see Fig.~\ref{exdip}). A
30-fold increase of the molecular fluorescence upon the decrease of
the tip-sample distance from 600~nm to 100~nm provides the expected
signature of near-field excitation. Furthermore, a careful scrutiny
of $I_{23}$ also reveals an oscillation as displayed by the zoom in
the inset. As indicated by the upper inset in Fig.~\ref{fdtd}f, this
is caused by the interference between the part of the molecular
fluorescence that is directed toward the detector and another part
that is reflected from the aluminum tip. The location $z\sim 330~nm$
for the first minimum is in fair agreement with a simple ray optics
picture that considers a phase shift of $\pi$ at the tip. Again the
FDTD calculations in Fig.~\ref{fdtd}f agree well with the data.

Figures~\ref{fdtd}c and d show experimentally determined $V$ and
$\psi$ as a function of the tip-molecule separation. As the molecule
gets closer to the tip, it is excited more strongly and the
visibility is increased. However, since $V$ is proportional to the
field of the molecular emission and therefore to the field of the
excitation beam at the position of the molecule\cite{gerhardt03},
its growth is slower than the fluorescence signal displayed in
Figs.~\ref{fdtd}b and f. In addition, $V$ is inversely proportional
to the field of the laser at the detector\cite{gerhardt03} so that
it is affected by the behavior of $I_{bg}$. Finally, we note that
the reflection of the molecular emission from the tip manifests
itself also in oscillations of $V$ and $\psi$ which correlate with
those of $I_{23}$ and $I_{bg}$. Figures~\ref{fdtd}g and h display
the results of the FDTD calculations which reproduce all features of
the experimental data. We emphasize, however, that despite an
excellent semiquantitative agreement with the measurements,
predicting the absolute magnitudes of quantities such as phase and
visibility require more precise knowledge of the geometry than has
been possible in this work.

\small
\normalsize

\begin{figure}[htb]
\centerline{\includegraphics[width=7.5cm]{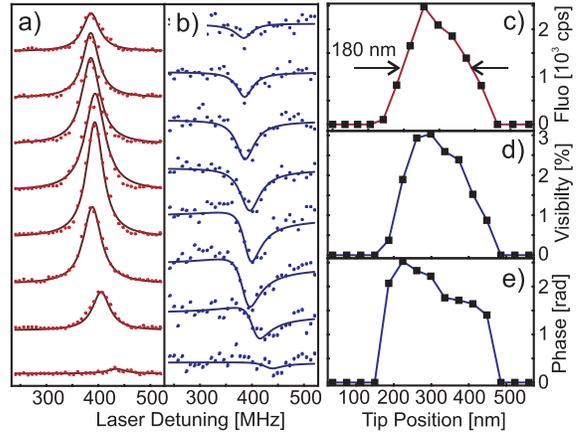}}
\caption{\label{xyscan} Fluorescence (a) and extinction (b) spectra
recorded during a lateral scan with a 100~nm aperture above a
molecule. Fluorescence peak intensity (c), visibility (d) and phase
(e) of the spectra from one line scan.}
\end{figure}

We have also investigated the spectra recorded at different lateral
tip-molecule displacements, while keeping the tip at an axial
distance of 90~nm from the sample. In Figs.~\ref{xyscan}a and b
examples of spectra on $PD_{23}$ and $PD_{21}$ from one line of a
600$\times$600~nm$^2$ lateral scan are presented. The line shape in
Fig.~\ref{xyscan}b is clearly modified within a displacement of a
fraction of a wavelength. Furthermore, the data reveal a small
residual position-dependent Stark shift. Figures~\ref{xyscan}c-e
plot $I_{23}$, $V$ and $\psi$ for various lateral displacements.

\small
\normalsize

In summary, we have presented direct near-field optical coherent
spectroscopy on single molecules without the need for any noise
suppression technique such as lock-in detection. We have
investigated the tip position dependence of both inelastic and
elastic components of the single molecule emission. Our FDTD
calculations show very good agreement with the experimental results
and reproduce all their central features.

We thank A. Renn and C. Hettich for fruitful discussions. This work
was financed by the Schweizerische Nationalfond (SNF) and the ETH
Zurich initiative on Quantum Systems and Information Technology
(QSIT). V. Sandoghdar's email address is vahid.sandoghdar@ethz.ch.
\\ $^*$ Present address:\ Abteilung Quanten-Informations-Verarbeitung, Universit\"at Ulm, D-89069 Ulm, Germany.


\end{document}